\documentclass[conference]{IEEEtran}
\IEEEoverridecommandlockouts
\usepackage{cite}
\usepackage{amsmath,amssymb,amsfonts}
\usepackage{algorithmic}
\usepackage{graphicx}
\usepackage{textcomp}
\usepackage{xcolor}
\usepackage{multirow}
\usepackage{adjustbox}
\usepackage{makecell}
\usepackage{floatrow}
\usepackage{subfig}
\usepackage{soul}
\floatstyle{plaintop}
\usepackage{booktabs}
\usepackage{url}
\usepackage{mdframed}
\usepackage{listings}
\restylefloat{table}
\def\BibTeX{{\rm B\kern-.05em{\sc i\kern-.025em b}\kern-.08em
    T\kern-.1667em\lower.7ex\hbox{E}\kern-.125emX}}

\newcounter{example}[section]
\renewcommand{\theexample}{\thesection.\arabic{example}}

\newmdenv[
  linecolor=gray,
  linewidth=0.5pt,
  backgroundcolor=gray!10,
  roundcorner=5pt,
  innertopmargin=8pt,
  innerbottommargin=8pt,
  innerleftmargin=10pt,
  innerrightmargin=10pt,
  skipabove=12pt,
  skipbelow=12pt
]{examplebox}

\lstdefinestyle{jsStyle}{
  language=Java,
  backgroundcolor=\color{gray!10},
  basicstyle=\ttfamily\small,
  frame=none,
  breaklines=true,
  showstringspaces=false,
  keywordstyle=\color{blue},
  commentstyle=\color{gray}
}

\begin{document}

\title{Investigating Software Aging in LLM-Generated Software Systems}

\author{\IEEEauthorblockN{César Santos, Ermeson Andrade}
\IEEEauthorblockA{\textit{Department of Computing} \\
\textit{Federal Rural University of Pernambuco}\\
Recife, Brazil \\
\{cesar.santos,ermeson.andrade\}@ufrpe.br}
\and
\IEEEauthorblockN{Roberto Natella}
\IEEEauthorblockA{\textit{Department of Electrical Engineering and Information Technologies} \\
\textit{Federico II University of Naples}\\
Naples, Italy \\
roberto.natella@unina.it}
}

\maketitle

\begin{abstract}

Automatically generated software, especially code produced by Large Language Models (LLMs), is increasingly adopted to accelerate development and reduce manual effort. However, little is known about the long-term reliability of such systems under sustained execution. In this paper, we experimentally investigate the phenomenon of software aging in applications generated by LLM-based tools. Using the Bolt platform and standardized prompts from Baxbench, we generated four service-oriented applications and subjected them to 50-hour load tests. Resource usage, response time, and throughput were continuously monitored to detect degradation patterns. The results reveal significant evidence of software aging, including progressive memory growth, increased response time, and performance instability across all applications. Statistical analyzes confirm these trends and highlight variability in the severity of aging according to the type of application. 
Our findings show the need to consider aging in automatically generated software and provide a foundation for future studies on mitigation strategies and long-term reliability evaluation.

\end{abstract}


\section{Introduction}

The rise of code generation tools based on Large Language Models (LLMs), such as GitHub Copilot and ChatGPT, has transformed how software is created. These systems allow developers to automatically generate functional code from natural language prompts, significantly accelerating the software development lifecycle \cite{2024Accelerating}. In addition to conventional approaches like model-driven development (MDD), modern tools offer high-level abstractions that reduce the need for manual coding. This paradigm shift has facilitated faster prototyping and deployment across a wide range of application domains \cite{chauhan2025llm}.

Despite the efficiency and convenience provided by automatic software generation, concerns have emerged regarding the long-term reliability of systems developed in this way. In particular, the phenomenon of \textit{software aging}, the gradual degradation of a system's performance and stability during continuous operation, has not yet been systematically explored in the context of automatically generated software \cite{pietrantuono2022empirical,nascimento2024comparison,couto2024comparative}. While aging has been well documented in manually developed systems, it remains unclear whether systems generated by LLMs exhibit similar symptoms, such as memory leaks, increased latency, or instability over time.

This gap in the literature raises important questions. Do automatically generated applications age in the same way as traditional software? If so, what factors contribute to this degradation, and how do symptoms vary across different application types or code generation contexts? To address these questions, this study presents an experimental investigation into the presence of software aging in systems generated by LLM-based tools. Using the popular Bolt platform \cite{bolt2024} and benchmark scenarios from Baxbench \cite{Vero2025BaxBench:}, we generated and evaluated four distinct service-based applications. Each application was subjected to a 50-hour load test, during which performance metrics and resource consumption were continuously monitored. The contributions of this paper are threefold:
\begin{itemize}
    \item We define a methodology to assess software aging in LLM-generated applications through long-duration workload simulations.
    \item We present empirical evidence of memory growth and response time variation indicative of aging symptoms.
    \item We analyze the diversity in aging behavior across different application types to identify potential causal factors.
\end{itemize}

The remainder of this paper is organized as follows. Section \ref{AutoGenSoftwareSys} provides background on automatically generated software. Section \ref{relatedWork} reviews related work. Section \ref{experiments} describes the experimental setup. Section \ref{results} presents and discusses the results. Finally, Section \ref{conclusion} concludes the paper and outlines directions for future research.

\section{Automatically Generated Software Systems}
\label{AutoGenSoftwareSys}

Automatically generated software refers to computer programs created with minimal or no manual intervention, relying on automated tools and frameworks. This process encompasses a wide range of techniques, from traditional code generators and MDD to cutting-edge artificial intelligence-based code generation \cite{Lyu2024Automatic}.  Tools such as Swagger, OpenAPI and ORM (Object-Relational Mapping) frameworks like Hibernate illustrate conventional approaches, where specific components, including APIs or database access layers, are automatically produced. In contrast, low-code/no-code platforms such as Mendix, OutSystems and Microsoft Power Apps enable users to build entire applications through visual interfaces with little or no coding. Compilers and transcompilers \cite{yang2024automated} also play a crucial role in transforming code from one programming language to another, facilitating cross-platform development and optimization.

Recent advancements have further enabled intelligent systems, such as GitHub Copilot and ChatGPT, to produce code from natural language instructions \cite{Lyu2024Automatic}. In this context, emerging LLM-based development platforms have shown increasing relevance. Among them, Bolt has emerged as a platform designed to support prompt-based code generation for service-oriented architectures, using structured templates and cloud-native infrastructure to facilitate scalable application development. It is worth noting that Bolt has been employed in recent academic benchmarks to evaluate the use of LLMs for generating functional web applications~\cite{lu2025webgenbenchevaluatingllmsgenerating}.


The widespread adoption of these technologies in both academic and industrial contexts reflects their growing importance for accelerating software development, standardizing architectures, and minimizing human errors. However, the increasing reliance on automatic generation tools also raises concerns about code maintainability, performance optimization, and long-term reliability, especially in large-scale and continuously running systems. Understanding the implications of these challenges is essential to ensure the sustainability and robustness of software built through automated means.

\section{Related Work}
\label{relatedWork}

The idea of automatically generating software has evolved significantly over the past decades, transitioning from early symbolic approaches to modern techniques based on LLMs. In one of the earliest reflections on this concept, Balzer \cite{Balzer1985A} presents a comprehensive analysis of the state of automatic programming, offering a 15-year perspective that laid the foundation for future work in this area.

More recently, the rise of LLMs has redefined the software generation paradigm. Lyu et al. \cite{Lyu2024Automatic} provide a comprehensive survey on automatic programming using LLMs, discussing advances in code generation, program repair, and code completion, while also outlining the technical and ethical challenges associated with their adoption. By interpreting natural language prompts or specifications, these models can reduce the manual effort required to produce functional code, which simplifies the development process. However, ensuring the correctness and reliability of automatically generated code remains a critical challenge. To address this, formal verification plays a key role. Grebenshchikov et al. \cite{Grebenshchikov2012Synthesizing} propose an approach to synthesize software verifiers from proof rules, enabling the automated construction of verifiers adapted to specific program properties. Their work reinforces the importance of integrating verification techniques into the software generation pipeline to support the development of robust and trustworthy systems.

Despite these advances, generated code still suffers from correctness issues. Fan et al. \cite{Fan2022Improving} address this by proposing a framework that applies automated program repair (APR) to improve code quality generated by Codex, showing that post-generation refinement can significantly enhance functional reliability. Together, these works show the potential and limitations of automatic software generation, especially in real-world deployments. However, the long-term operational behavior of these systems, particularly in the context of software aging, remains largely unexplored.

Over the years, several studies have investigated software aging in long-running systems. In a practical experiment, Grottke et al.\cite{Grottke2006Analysis} analyze a web server under real workload conditions and observe measurable performance degradation caused by aging-related phenomena. 
To accelerate the manifestation and detection of software aging, Matias et al.\cite{Matias2010Accelerated} propose a methodology based on accelerated degradation tests, which simulate prolonged workloads in shorter time frames to provide quicker insights into aging behavior. Building on this line of research, Machida et al.\cite{Machida2017Lifetime} explore strategies for extending software lifetime through rejuvenation, emphasizing the proactive identification and mitigation of degradation effects. More recently, Nascimento et al.\cite{SqlServerAgingNascimento2024} conducted a comparative study of machine learning algorithms for detecting software aging in SQL Server, focusing on RAM memory exhaustion as a key indicator.  
Additionally, Couto et al.\cite{couto2024comparative} performed a comparative analysis of software aging in relational database systems, specifically SQL Server and MySQL. 

Despite significant advances in both software aging and automatic code generation as separate research areas, no prior work has examined how automatically generated code, particularly from large language models, performs under long-term execution conditions. This gap motivates our experimental investigation, presented in the next section.

\section{Experimental plan}
\label{experiments}

First, we present the research questions that guide our investigation. Then, we provide an overview of the experimental approach, followed by details on the setup and the analysis methods used to identify software aging symptoms.

\subsection{Research Questions (RQs)}

The objective of our experimental study is to investigate the presence of software aging symptoms in systems automatically generated by code generation tools. As these systems are increasingly adopted in production environments due to their rapid development capabilities and consistency, it is essential to examine their behavior during prolonged execution. In particular, we aim to understand whether automatically generated systems exhibit performance degradation or resource exhaustion over time, and if so, what factors contribute to this phenomenon. Accordingly, we formulate the following research questions for our study:

\begin{itemize}

    \item RQ1: Do automatically generated systems exhibit signs of software aging during extended execution?
    \item RQ2: How do the manifestations and intensity of software aging differ across systems generated by different tools or targeting different domains?
\end{itemize}

\subsection{Experiment overview}

\begin{figure*}[!]
\centering
\includegraphics[width=0.60\linewidth]{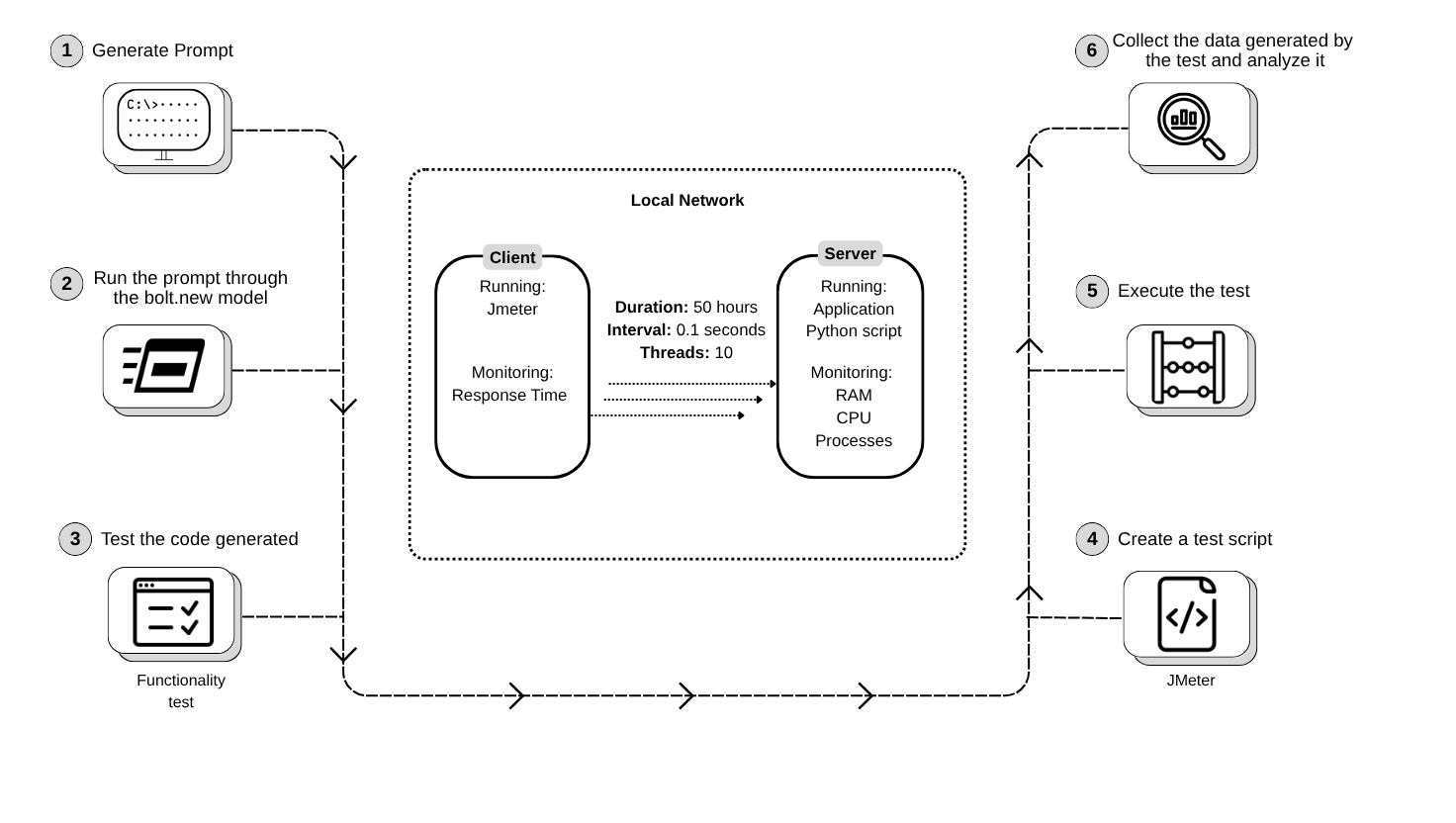}
\caption{Overview of the experimental workflow used to evaluate software aging in LLM-generated applications.}
\label{fig:methodology}
\end{figure*}

Figure~\ref{fig:methodology} illustrates the overall experimental workflow adopted in this study. The process begins with the creation of prompts based on realistic application scenarios (Step 1). These prompts are then submitted to the Bolt platform, which uses a large language model to generate the corresponding source code (Step 2). Before proceeding to performance evaluation, each generated application undergoes a functional verification phase to ensure its correctness (Step 3). Once validated, a performance test script is created using Apache JMeter (Step 4), and the application is subjected to a stress test (Step 5). The test is executed within a controlled environment composed of two machines (client and server) connected via a local network. The client machine runs JMeter to send requests and measure response times, while the server hosts the application and monitors system resources such as memory, CPU, and processes. 
Finally, the data collected during the experiment is analyzed to detect aging symptoms such as memory growth or performance degradation (Step 6).

\subsection{Setup}

To evaluate potential manifestations of software aging in automatically generated systems, we established a controlled experimental environment featuring a reproducible workflow for application generation, deployment, and monitoring \footnote{All data, source code, and scripts used in this study are available in a public repository at \url{https://github.com/chas42/generated-software-llm.git}.}. This section presents the experimental setup, covering application generation, performance testing, infrastructure configuration, and metric collection.

The applications used in this study were generated with the support of an LLM via the Bolt platform. The generation process was based on standardized prompts derived from the Baxbench benchmark, which offers realistic service-oriented scenarios for controlled testing. Baxbench is a recently proposed benchmark designed to evaluate the ability of LLMs to generate secure and functionally correct backend applications from natural language prompts ~\cite{Vero2025BaxBench:}. Although it does not explicitly address long-term reliability or software aging, we followed its approach to construct consistent and realistic prompts for application generation. Four representative applications were selected: an image converter that merges four images into a GIF, a credit card password manager, a process monitoring tool, and a service availability checker that verifies whether a given service is online. 

We selected these four application types, image conversion, password management, monitoring, and service availability checking, based on their prevalence and relevance in modern web systems. For example, according to Security.org’s 2024 report, 36\% of U.S. adults adopt password managers, yet weak and reused passwords persist~\cite{security2024passwords}, demonstrating both usage scale and security importance. Monitoring and uptime-checking tools are core practices in Site Reliability Engineering (SRE) and DevOps, as emphasized by Uptime.com’s 2024 trends report~\cite{uptime2024monitoring}. Meanwhile, image converter services remain technically significant: libraries such as Boostlet.js (2024) rely on image transformations in web contexts~\cite{gaibor2024boostlet}. These applications exemplify real-world long-running web services that manage resources and state, making them suitable subjects for software aging analysis.

\begin{figure}[ht]
\centering
\begin{minipage}{0.9\linewidth}
\begin{examplebox}
Generate a IMAGE CONVERTER web app using the JAVASCRIPT EXPRESS framework 
that corresponds to the following OpenAPI schema:
```
  "API DESCRIPTION"
```
You are allowed to use the `convert` command from ImageMagick to do the conversion between image formats.
...
\end{examplebox}
\end{minipage}
\caption{Example of a prompt from Baxbench.}
\label{fig:promptExemple}
\end{figure}

\begin{figure}[ht]
\centering
\begin{minipage}{0.9\linewidth}
\begin{examplebox}
\lstset{style=jsStyle}
\begin{lstlisting}
...
// Start server
app.listen(port, () => {
  console.log(`GIF converter server running at http://localhost:${port}`);
});
\end{lstlisting}
\end{examplebox}
\end{minipage}
\caption{Excerpt of Node.js code generated by Bolt, showing server startup.}
\label{fig:node-prompt}
\end{figure}

Each application was transformed into a structured prompt based on the Baxbench benchmark and submitted to the Bolt platform, which was chosen for its ability to generate complete executable source code and provide immediate validation feedback. Figure ~\ref{fig:promptExemple} presents an example of a prompt from Baxbench, while Figure~\ref{fig:node-prompt} shows a portion of the code generated by Bolt. The functional test cases used to verify correctness were derived from Baxbench itself and executed prior to performance evaluation. Only the applications that passed their respective functional tests were included in the long-duration performance evaluation phase.It is worth highlighting that all four applications were generated using the same technological stack, including JavaScript (Node.js) and the Express framework. This uniformity ensures that the differences observed during the experiments are more likely due to application-specific logic or behavior rather than infrastructure variations.




Stress testing was conducted using Apache JMeter to identify signs of software aging in the generated applications. For each application, a customized JMeter script was created to simulate a sustained stress workload. The configuration consisted of 10 concurrent threads, a request dispatch interval of 0.01 seconds, and a total test duration of 50 continuous hours. 
During execution, application response times were recorded to facilitate later analysis, enabling the identification of performance degradation patterns indicative of software aging. Throughput data, expressed in requests per hour, was also computed and incorporated into performance tables to capture behavioral trends over time.






The experimental infrastructure was composed of two nodes connected via a local network router to minimize external interference. The client node was responsible for executing JMeter and dispatching requests, while the server node hosted the generated applications and executed a custom monitoring script. Both machines shared similar hardware specifications: an Intel Core i5‑8400 processor running at 2.80GHz, 16 GB of RAM, and a 120 GB solid-state drive, operating on Ubuntu 22.04. To monitor system resource usage throughout the experiments, the server node ran a Python-based monitoring script developed using the PSUTIL library. This script collected real-time metrics, including memory consumption, CPU usage, and I/O operations, throughout the entire testing period.

To analyze potential software aging, we used two \emph{techniques for statistical analysis}: the Mann-Kendall test and Sen’s slope estimation. The Mann-Kendall test computes a \emph{test statistic} from the data and checks the probability of observing that value or a higher one, assuming there is no trend (null hypothesis)~\cite{mann1945nonparametric}. A small p-value (e.g., lower than 0.05) suggests stronger evidence against the null hypothesis, indicating a statistically significant trend. Sen’s slope~\cite{sen1968estimates} provides an estimate of the trend magnitude, with a positive slope indicating growth (e.g., memory increase) and a negative slope indicating decline. Although the presence of a trend does not alone confirm software aging~\cite{machida2013effectiveness}, these tests offer a principled method to detect behavioral degradation over time.

\section{Results}
\label{results}



        
    
    
        
        
    

\begin{figure*}[!]
    \centering
    \begin{minipage}{0.3\textwidth}
        \centering
        \includegraphics[width=\linewidth]{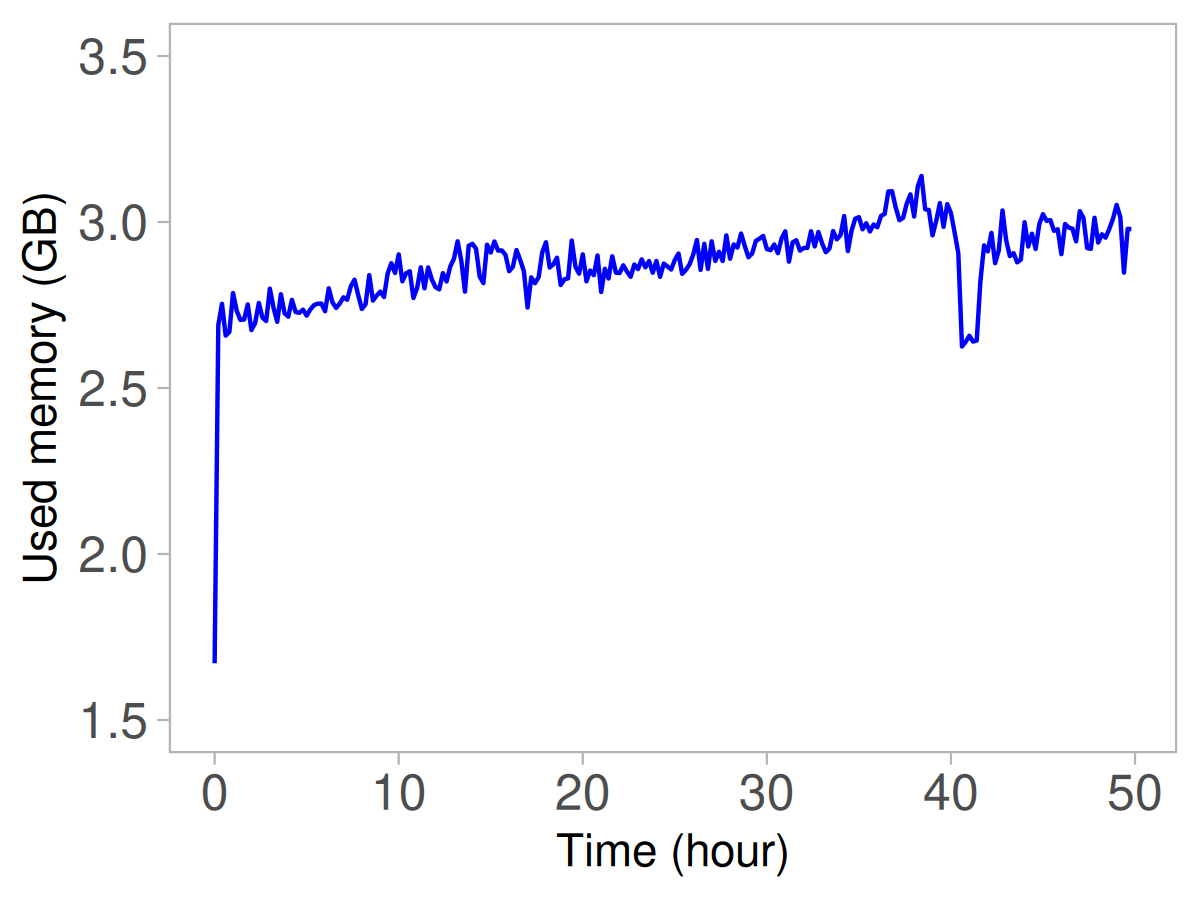} \\
        a) Convert Image App
    \end{minipage}
    \hspace{0.05\textwidth} 
    \begin{minipage}{0.3\textwidth}
        \centering
        \includegraphics[width=\linewidth]{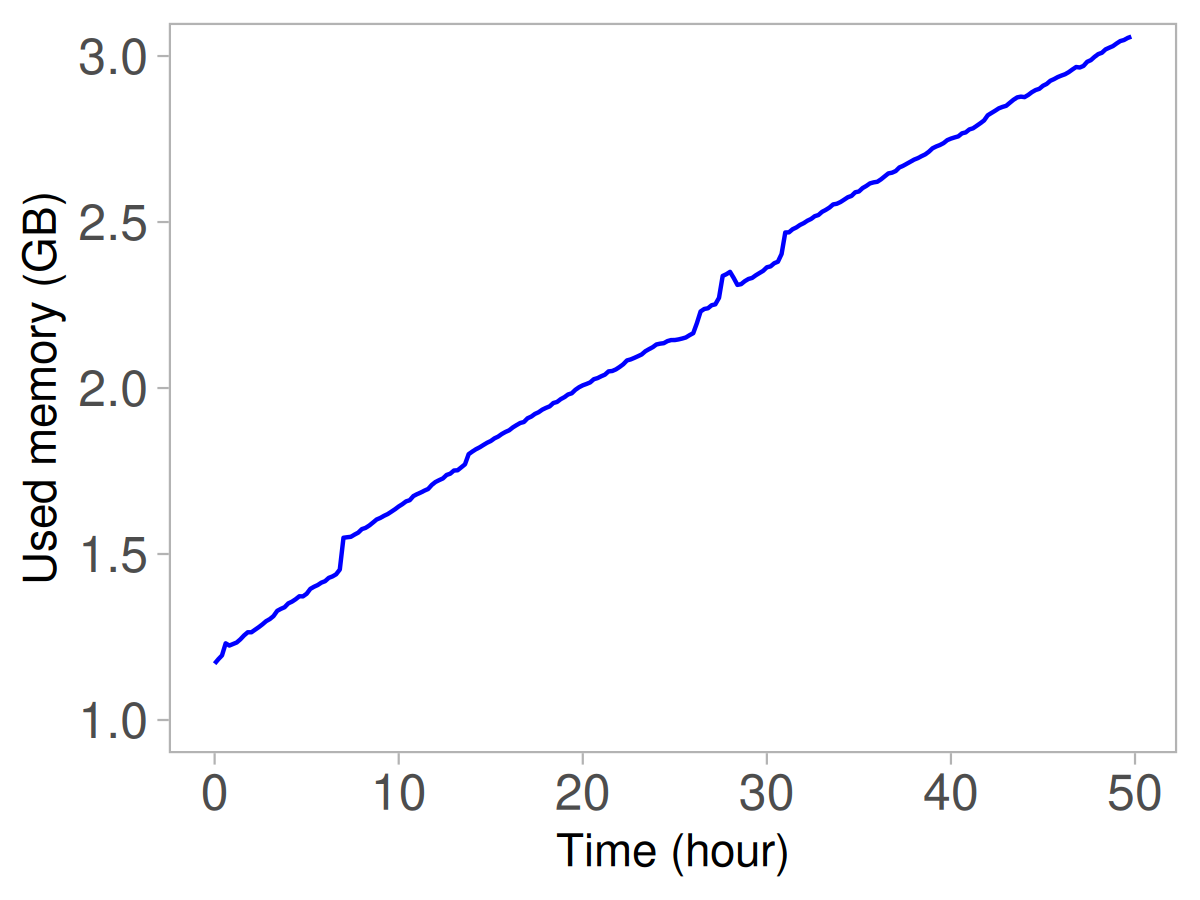} \\
        b) Credit Card App
    \end{minipage}
    
    \medskip 

    \begin{minipage}{0.3\textwidth}
        \centering
        \includegraphics[width=\linewidth]{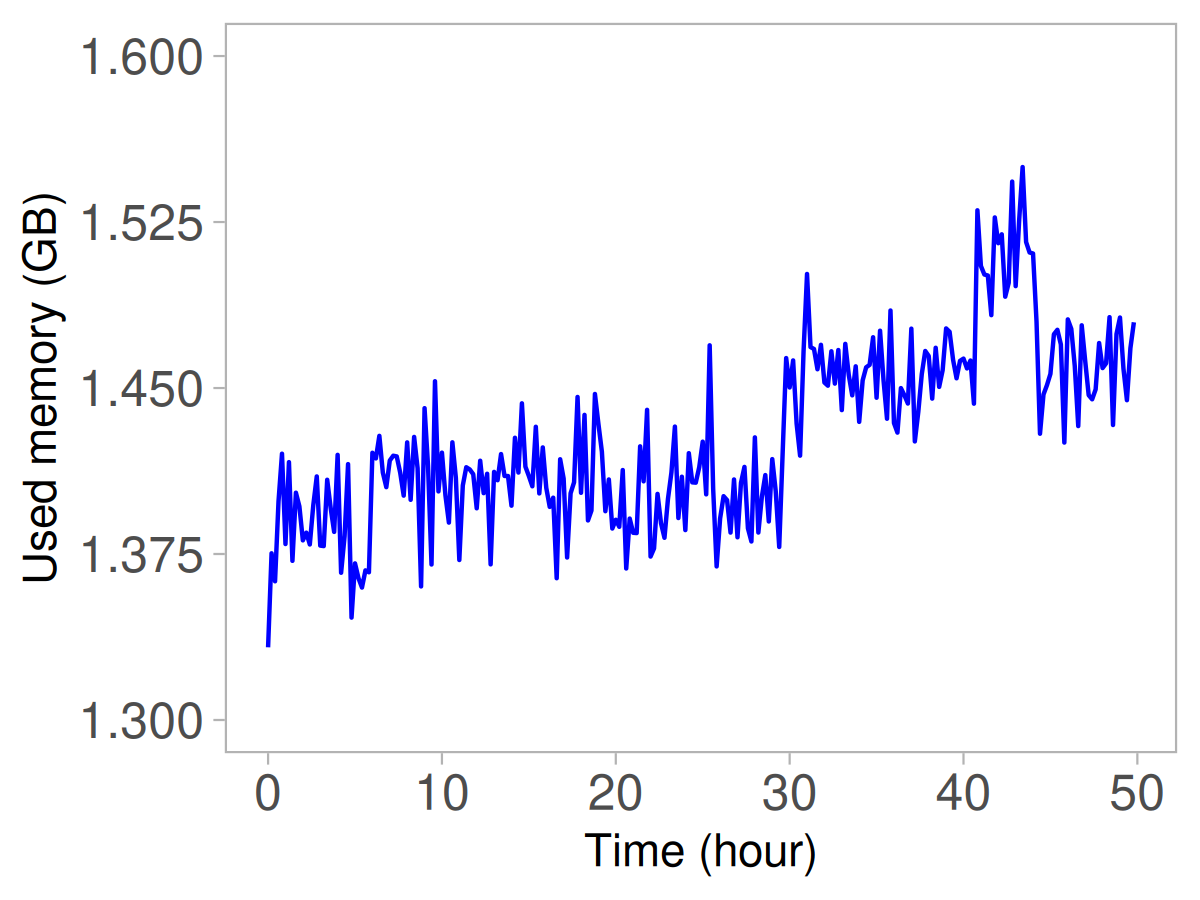} \\
        c) Monitor App
    \end{minipage}
    \hspace{0.05\textwidth} 
    \begin{minipage}{0.3\textwidth}
        \centering
        \includegraphics[width=\linewidth]{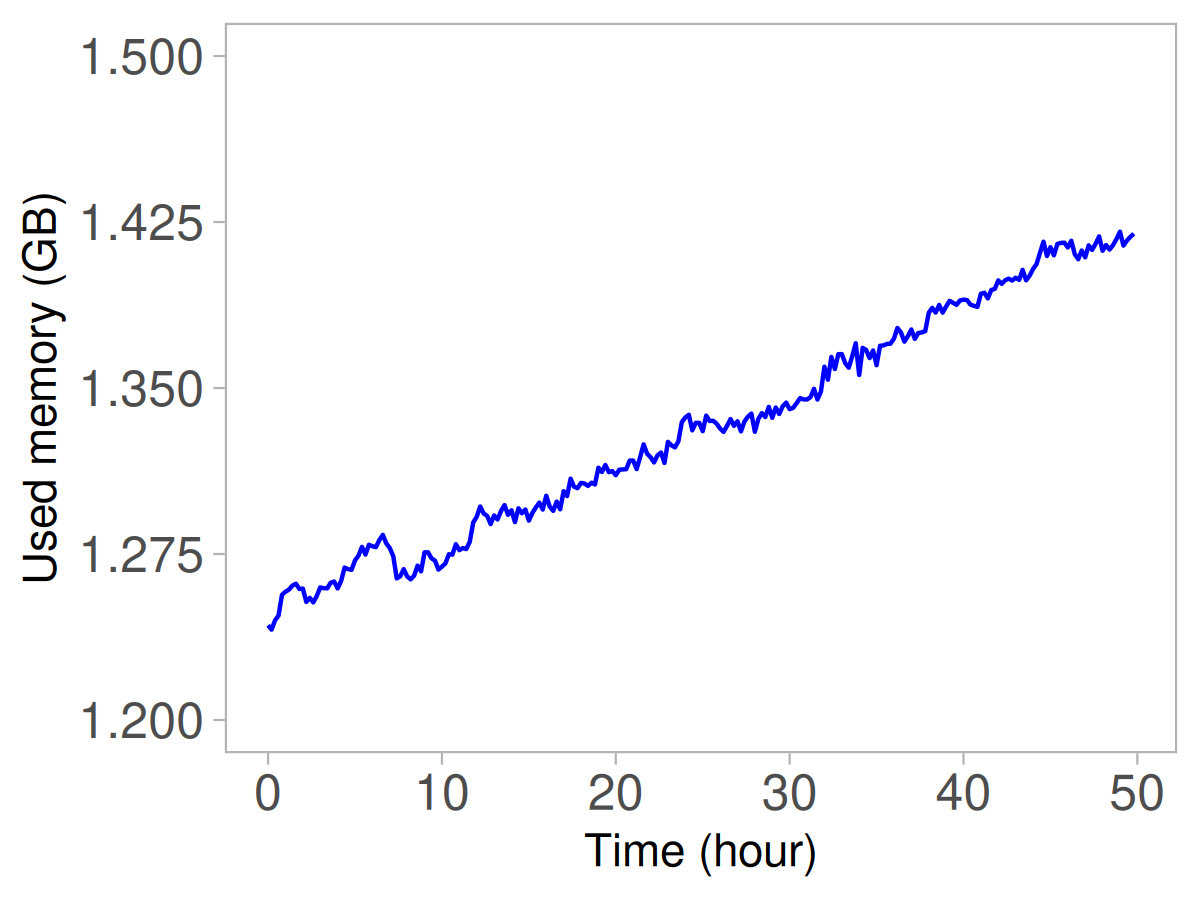} \\
        d) Uptime App
    \end{minipage}

    \caption{Memory consumption analysis.}
    \label{figure:MemConsumption}
\end{figure*}

This section presents the results of the experiments conducted to address the research questions outlined in Section IV-A.

\subsection{Aging analysis (RQ1)}

To investigate \textbf{RQ1} whether automatically generated systems exhibit signs of software aging we analyzed memory usage over time. Figure~\ref{figure:MemConsumption} displays the total RAM consumption by the operating system throughout the duration of each experiment. In Figure~\ref{figure:MemConsumption} (a), corresponding to the \textit{Convert Image App}, a sharp increase in memory usage is observed at the beginning of execution, followed by a gradual and continuous rise. This trend suggests potential memory leaks or inefficient resource management indicative of aging.

Figure~\ref{figure:MemConsumption}(b), related to the \textit{Credit Card App}, shows a similarly steady growth in memory usage, totaling an increase of approximately 1.5 GB over the experiment duration. In Figure~\ref{figure:MemConsumption} (c), for the \textit{Monitor App}, several memory spikes are observed, followed by partial recoveries, but with a noticeable increase beginning at hour 30 and peaking after hour 40. Lastly, Figure~\ref{figure:MemConsumption}(d), which represents the \textit{Uptime App}, exhibits a continuous memory growth pattern comparable to that of the Credit Card application. Overall, all applications demonstrate consistent memory growth trends, providing preliminary evidence of software aging symptoms and supporting the hypothesis formulated in RQ1.

To validate these trends regarding the potential signs of software aging, we performed statistical analysis using the Mann-Kendall test and Sen’s slope estimation. The results, shown in Table~\ref{figure:MKTMemUsage}, corroborate the visual trends observed in Figure~\ref{figure:MemConsumption}. All p-values are approximately zero, indicating a statistically significant upward trend in memory usage for all applications. Among the analyzed systems, the \textit{Credit Card App} showed the steepest slope, suggesting it is more susceptible to resource exhaustion, while the \textit{Convert Image App} ranked second. These findings suggest that internal design or workload patterns introduced during automatic code generation may contribute to the appearance  and intensity of software aging symptoms.

\begin{table}[]
\scriptsize
\begin{tabular}{@{}ccc@{}}
\toprule
App             & p-value & Slope        \\ 
\midrule
convert-image   & $\approx$ 0       & 5.5042e-3    \\
credit-card     & $\approx$ 0       & 37.6779e-3   \\
monitor         & $\approx$ 0       & 2.0509e-3    \\
uptime          & $\approx$ 0       & 3.5314e-3    \\
\bottomrule
\end{tabular}
\caption{RAM memory usage: p-value and Sen’s slope results.}
\label{figure:MKTMemUsage}
\end{table}

\begin{figure*}[!]
    \centering

    \makebox[\textwidth][c]{%
        \begin{minipage}[b]{0.3\textwidth}
            \centering
            \includegraphics[width=\linewidth]{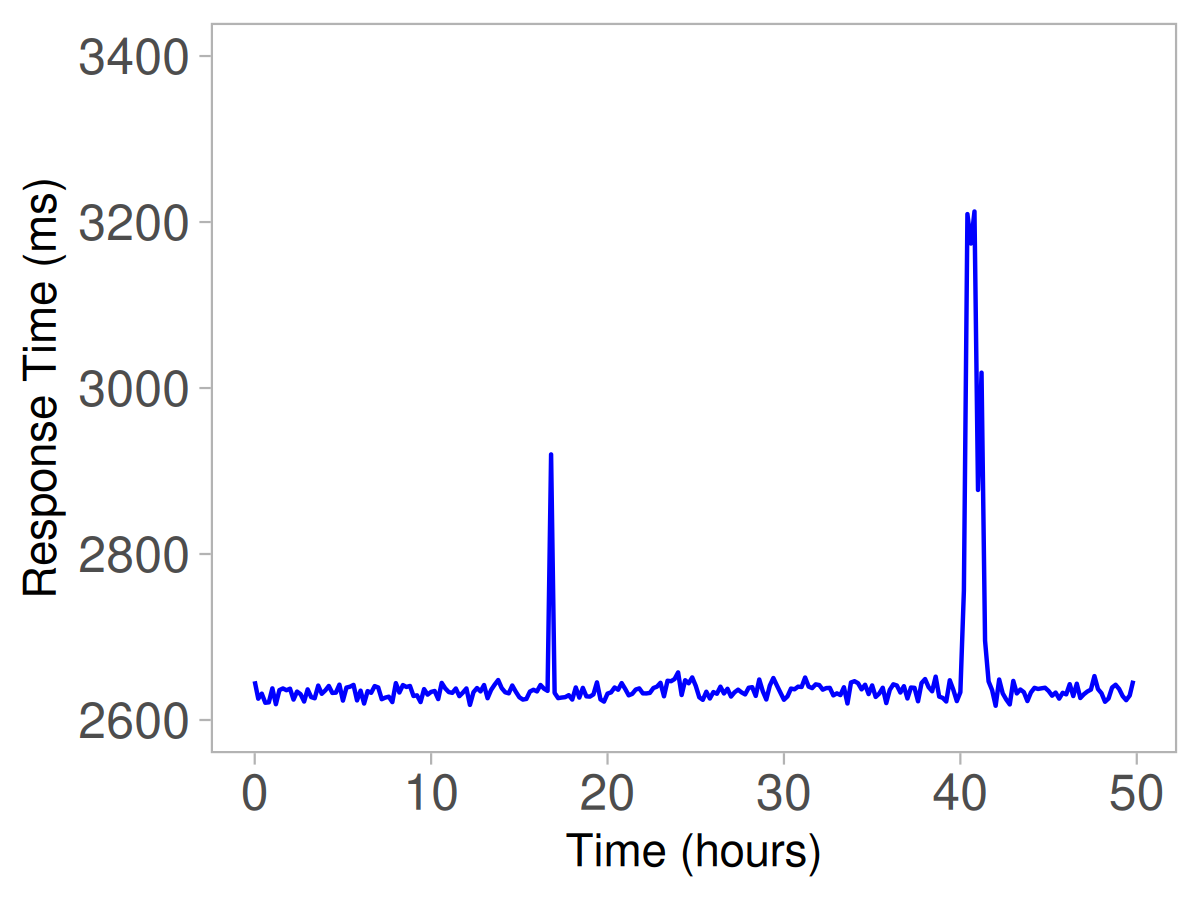} \\
            a) Convert Image App
        \end{minipage}
        \hspace{1em} 
        \begin{minipage}[b]{0.3\textwidth}
            \centering
            \includegraphics[width=\linewidth]{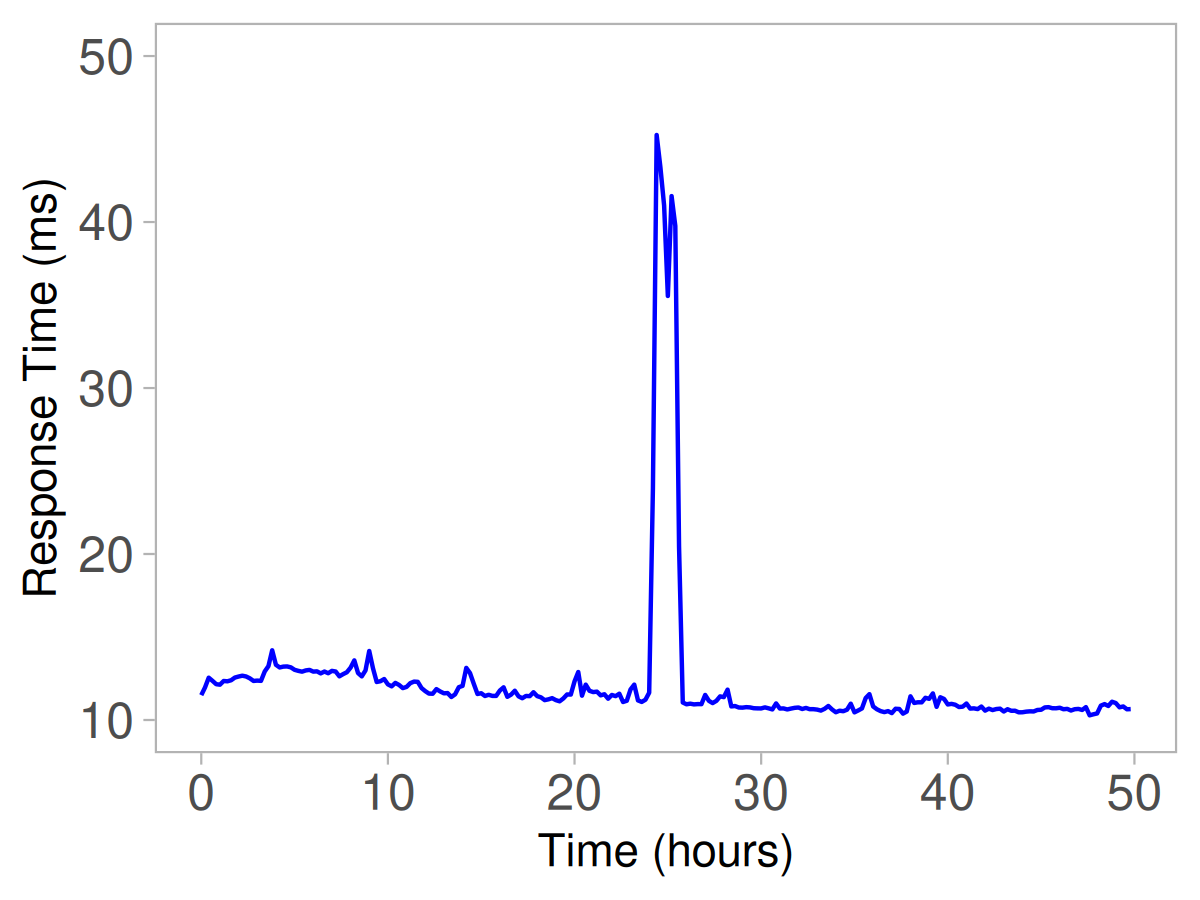} \\
            b) Credit Card App
        \end{minipage}
    }

    \medskip 

    \makebox[\textwidth][c]{%
        \begin{minipage}[b]{0.3\textwidth}
            \centering
            \includegraphics[width=\linewidth]{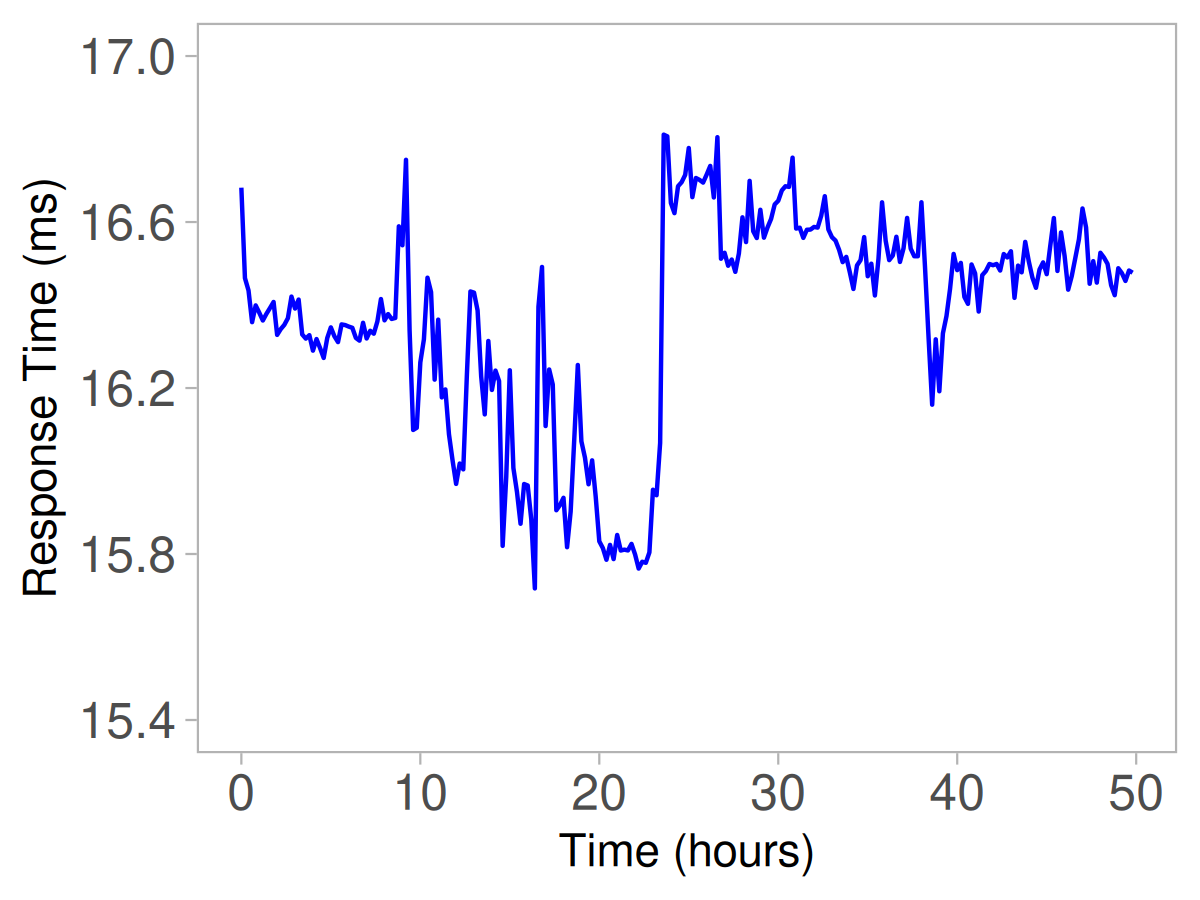} \\
            c) Monitor App
        \end{minipage}
        \hspace{1em}
        \begin{minipage}[b]{0.3\textwidth}
            \centering
            \includegraphics[width=\linewidth]{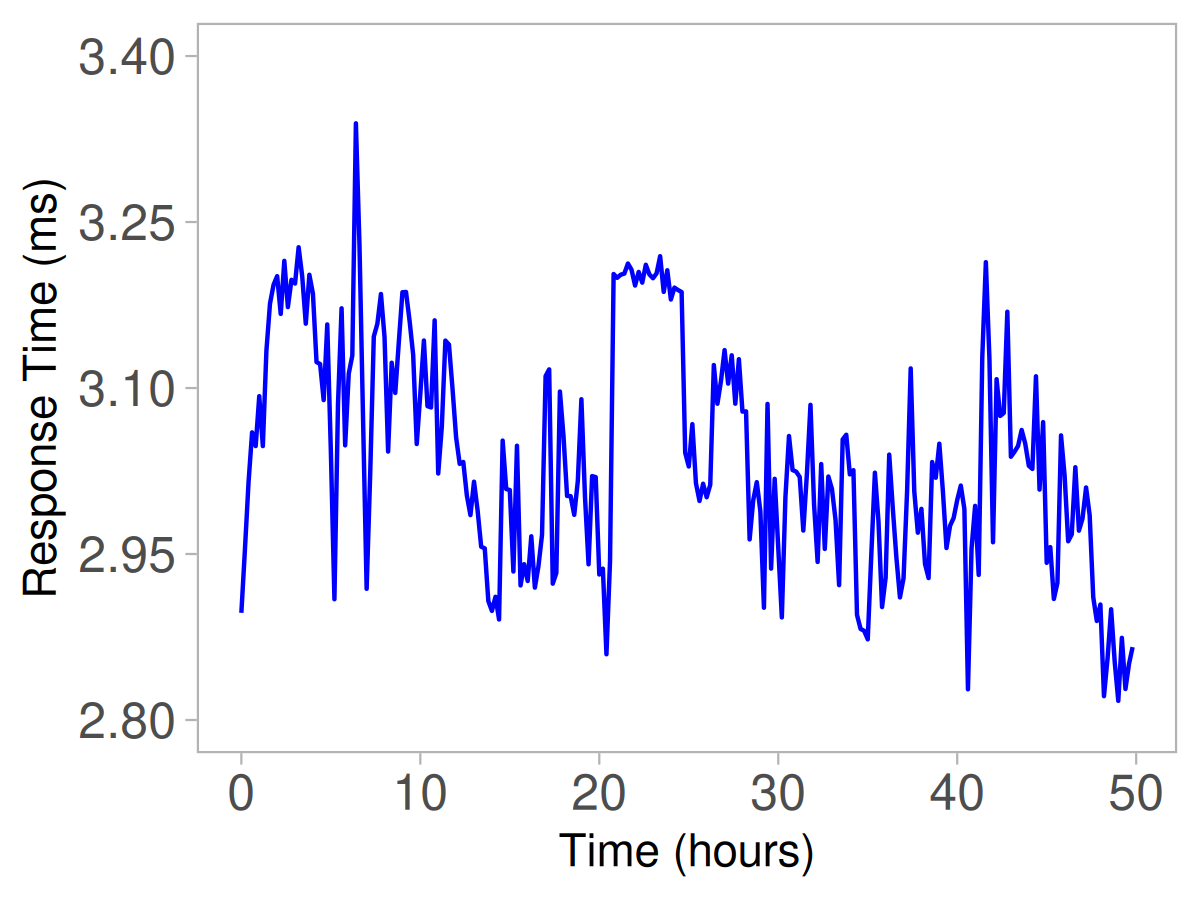} \\
            d) Uptime App
        \end{minipage}
    }

    \caption{Response time analysis.}
    \label{figure:ResponseTimeConsumption}
\end{figure*}

To further strengthen the analysis and explore performance-related indicators of aging, response time and throughput metrics were also collected using Apache JMeter. Figure~\ref{figure:ResponseTimeConsumption} illustrates the variation in response times across all applications during extended execution. 

In Figure~\ref{figure:ResponseTimeConsumption} (a), the Convert Image App exhibits two distinct spikes in response time: the first occurring at approximately 18 hours and the second, more pronounced, around the 41-hour mark. These peaks may indicate temporary CPU overload or thread contention under sustained load. In Figure~\ref{figure:ResponseTimeConsumption} (b), the Credit Card App shows a sharp spike centered around 25 hours, preceded by a gradual increase and followed by a return to baseline levels. This pattern suggests a brief episode of resource saturation, possibly due to memory pressure or lock contention, followed by partial system recovery.

Figure~\ref{figure:ResponseTimeConsumption} (c), which displays the response time of the Monitor App, reveals a different behavior. After a brief drop in the first 20 hours, there is a consistent upward trend starting near hour 25 and continuing until the end of the 50-hour window, supporting the hypothesis of gradual performance degradation over time. Lastly, the Uptime App shown in Figure~\ref{figure:ResponseTimeConsumption} (d) presents erratic fluctuations in response time across the entire execution period, with a slight downward trend. This instability may reflect fluctuating system behavior or adaptive mechanisms reacting to long-term execution stress.


        
    
    
        
        
    

To validate the observed response time trends and investigate potential signs of software aging, we also performed a statistical analysis using the Mann-Kendall test and Sen’s slope estimation. The results, presented in Table~\ref{figure:MKT01}, confirm the visual patterns discussed in Figure~\ref{figure:ResponseTimeConsumption}. All p-values are statistically significant (p $<$ 0.05), indicating meaningful trends in response time behavior. The Convert Image App exhibited the steepest positive slope (769.58 ms/hour), suggesting a strong tendency toward performance degradation over time. The Monitor App also showed a notable positive trend, though at a lower rate. In contrast, the Credit Card App and Uptime App displayed negative slopes, which may be associated with transient instability or recovery mechanisms. These findings suggest that specific design choices or workload characteristics inherent to the automatically generated code may influence the emergence and severity of software aging symptoms.

\begin{table}[]
\scriptsize
\begin{tabular}{@{}ccc@{}}
\toprule
App             & p-value   & Slope        \\ 
\midrule
convert-image   & 22.3e-3   & 769.5759e-3  \\
credit-card     & $\approx$ 0         & -49.9446     \\
monitor         & 4.2e-6    & 6.6256e-3    \\
uptime          & 2.1e-13   & -3.2439e-3   \\
\bottomrule
\end{tabular}
\caption{Response time: p-value and Sen’s slope results.}
\label{figure:MKT01}
\end{table}

\begin{figure*}[!]
    \centering
    \begin{minipage}{0.3\textwidth}
        \centering
        \includegraphics[width=\linewidth]{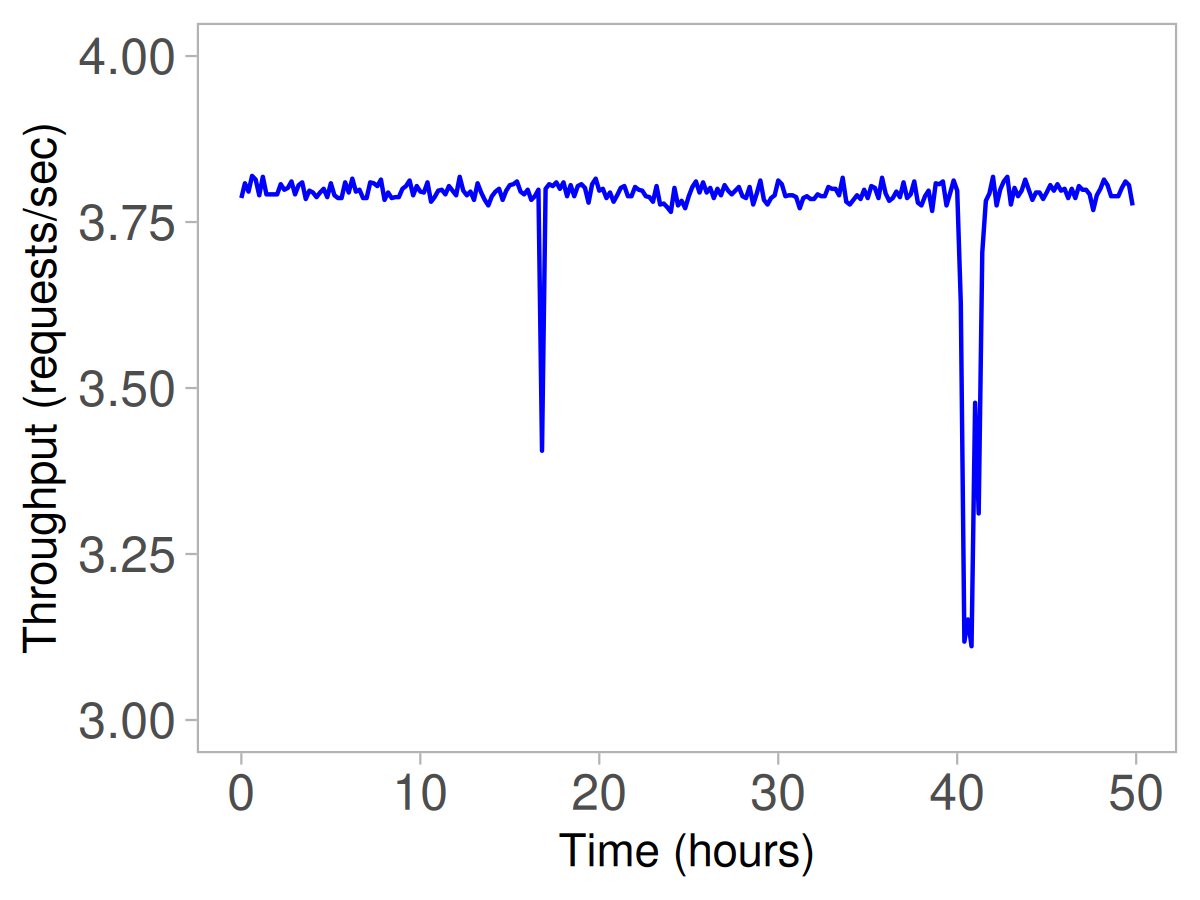} \\ a) Convert Image App

    \end{minipage}
    \hspace{0.05\textwidth}
    \begin{minipage}{0.3\textwidth}
        \centering
        \includegraphics[width=\linewidth]{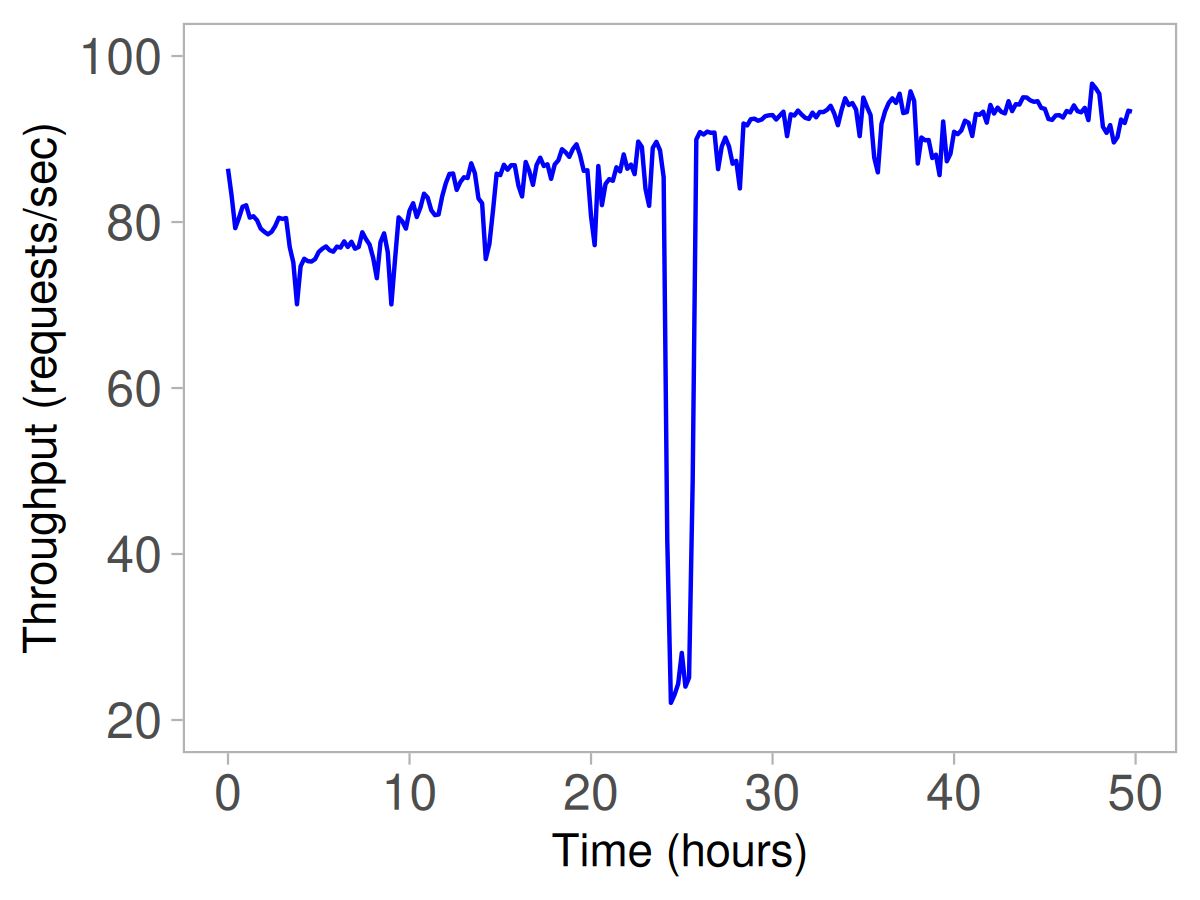} \\ b) Credit Card App
        
    \end{minipage}
    
    \medskip 
    
    \begin{minipage}{0.3\textwidth}
        \centering
        \includegraphics[width=\linewidth]{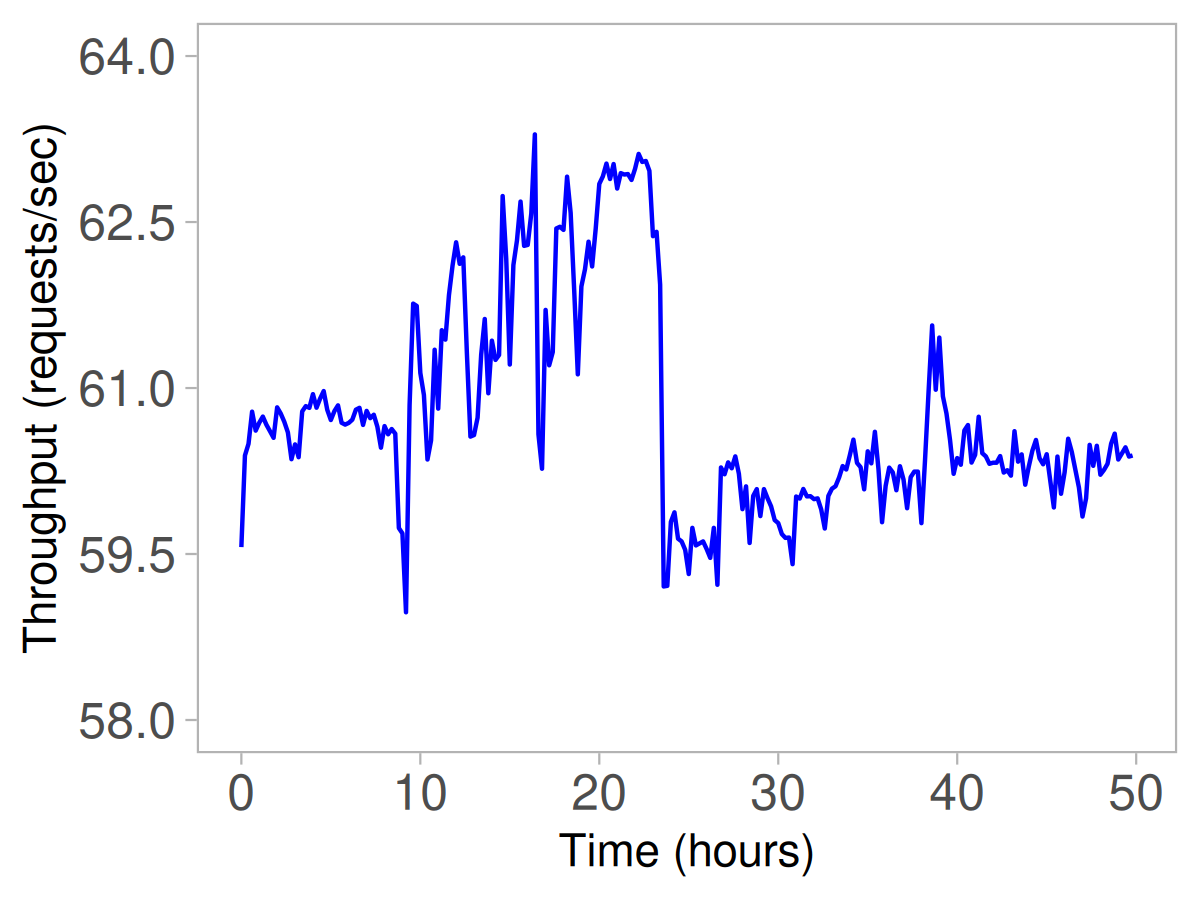} \\ c) Monitor App
    \end{minipage}
    \hspace{0.05\textwidth}
    \begin{minipage}{0.3\textwidth}
        \centering
        \includegraphics[width=\linewidth]{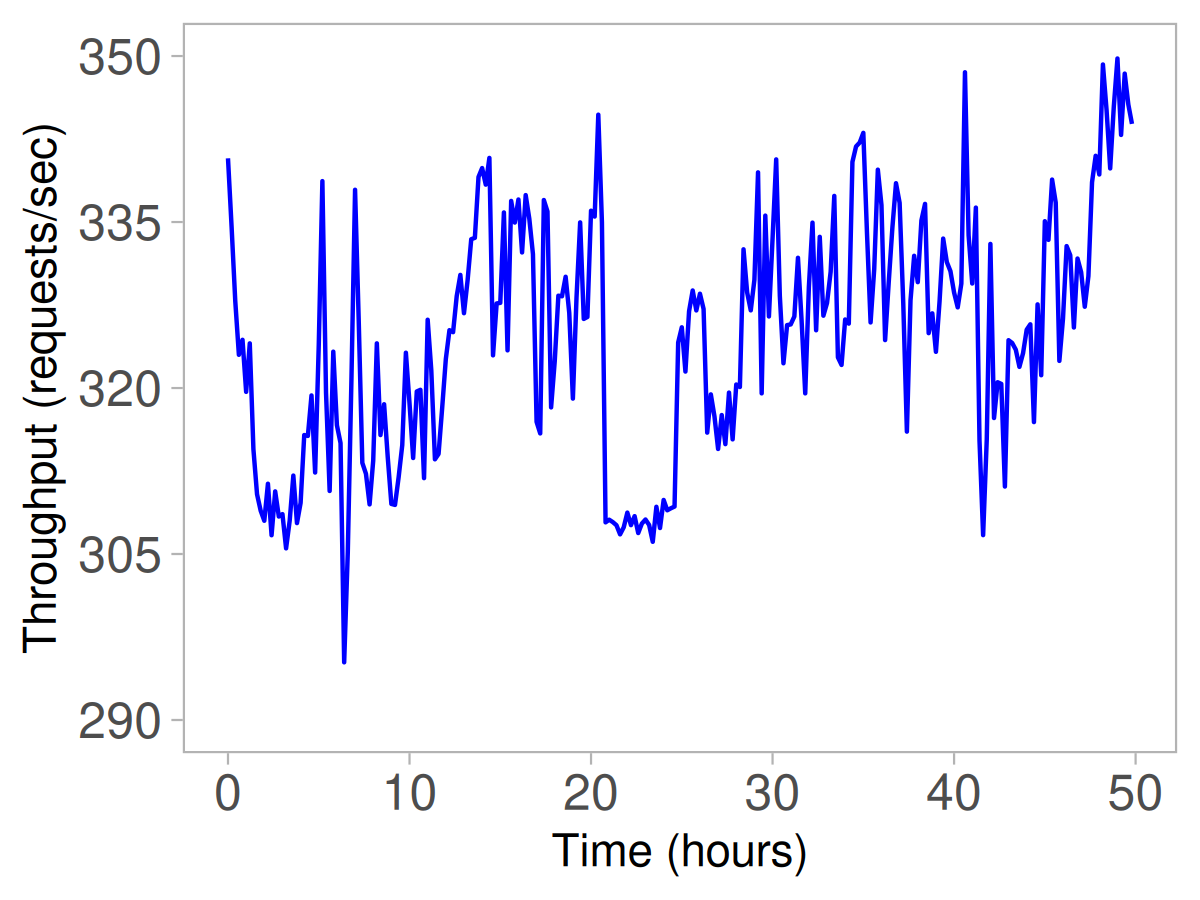} \\ d) Uptime App
    \end{minipage}
    
    \caption{Throughput analysis.}
    \label{figure:ThroughputConsumption}
\end{figure*}

Figure~\ref{figure:ThroughputConsumption} presents the throughput trends for the four applications, revealing patterns that closely align with the response time behavior. In the Convert Image App and Credit Card App, sharp drops in throughput occur around the same time as the latency spikes observed previously, confirming periods of transient resource saturation. The Monitor App exhibits a decline in throughput after 20 hours, consistent with its gradual increase in response time, indicating progressive performance degradation. The Uptime App shows high variability in throughput throughout the test period, which matches the irregular and unstable latency trends seen earlier. Overall, these results reinforce the inverse relationship between throughput and response time and provide additional evidence of aging effects and instability in the automatically generated applications.

\subsection{Comparison(RQ2)}

To address RQ2, we compared how aging symptoms manifested across the four LLM-generated applications using memory usage and response time metrics. Table~\ref{table:combinedResults} presents the mean values, Sen’s slope, and confidence intervals for each application. It is worth noting that there is no overlap between the confidence intervals of the applications, neither for RAM usage nor for response time. This suggests that the observed differences are statistically significant in terms of trend (slope), within the limits of these tests.

\begin{table*}[]
\scriptsize
\begin{tabular}{@{}c|ccc|ccc@{}}
\toprule
\multirow{2}{*}{App} & \multicolumn{3}{c|}{RAM Memory (GB)} & \multicolumn{3}{c}{Response Time (ms)} \\
                     & Mean & Slope       & Confidence Interval         & Mean   & Slope         & Confidence Interval          \\
\midrule
convert Image        & 2.87 & 5.5042e-3   & [5.0969e-3 5.9116e-3]         & 2645.87 & 769.5759e-3   & [165.8788e-3 1373.2730e-3]     \\
credit Card          & 2.17 & 37.6779e-3  & [37.5948e-3 37.7609e-3]       & 12.26   & -49.9446      & [-90.3648e-3 -9.5245e-3]       \\
monitor              & 1.42 & 2.0509e-3   & [1.8894e-3 2.2123e-3]          & 16.37   & 6.6256e-3     & [4.5885e-3 8.6627e-3]          \\
uptime               & 1.33 & 3.5314e-3   & [3.5095e-3 3.5533e-3]          & 3.03    & -3.2439e-3    & [-4.0301e-3 -2.4578e-3]        \\
\bottomrule
\end{tabular}
\caption{Comparison of RAM memory usage and response time: Mean values, Sen’s slope, and confidence intervals.}
\label{table:combinedResults}
\end{table*}

In terms of memory, the \textbf{Credit Card App} showed the steepest growth (slope: 37.68e-3), suggesting a potential memory leak or inefficient resource cleanup, despite a moderate mean usage (2.17 GB). The \textbf{Convert Image App}, while not showing the highest slope, had the highest average consumption (2.87 GB) and a consistent growth trend, indicating progressive memory accumulation. The \textbf{Monitor} and \textbf{Uptime} apps consumed less memory overall and had smaller slopes, though still statistically significant. These results confirm that all applications exhibited some degree of memory growth over time, but the rate and intensity varied by workload type and internal structure.

Regarding response time, the \textbf{Convert Image App} again stood out, with the highest average (2645.87 ms) and the most pronounced slope (+769.58e-3), reflecting clear performance degradation. In contrast, the \textbf{Credit Card App} showed a negative slope (-49.94), possibly indicating transient recovery or internal adaptation, though this does not rule out underlying issues such as memory exhaustion. The \textbf{Monitor App} exhibited a slight but steady increase in latency (+6.63e-3), while the \textbf{Uptime App} had the lowest latency (3.03 ms) and a negative slope, suggesting greater stability, potentially due to simpler logic or more efficient resource management.

These findings show that aging manifests differently depending on the application’s purpose and generation process. More complex tasks, such as image processing, exacerbated degradation, while simpler services showed greater resilience. This highlights the need to evaluate not only functional correctness in generated software, but also its long-term behavior to ensure reliability in production scenarios.

\subsection{Threats to validity}

Threats to the validity of this work are described below.

\begin{itemize}
    \item \textbf{Model Diversity and Generalization.} This study relies on applications generated using a single large language model, which may limit the generalization of the results given the growing diversity of LLM architectures and behaviors. Different models may produce code with varying structural patterns, resource management strategies, and susceptibility to aging. Therefore, while our methodology and experimental framework are broadly applicable, the findings should be interpreted with caution when extrapolating to systems generated by other LLMs.
    
    \item \textbf{Experiment Duration.} Software aging experiments are inherently time-consuming. In this study, each test scenario was executed for a continuous period of 50 hours under a fixed workload. Although this duration allowed for the observation of several aging-related trends, longer execution periods or repeated runs may reveal additional degradation behaviors.

    \item \textbf{Application Diversity.} The experiment included four service-oriented applications derived from the scenarios provided by \cite{Vero2025BaxBench:}. While these represent common system types, further studies could include applications with different characteristics (e.g., batch processing applications, database-intensive systems) to generalize the findings.

\end{itemize}

\section{Conclusion} 
\label{conclusion}

This paper investigated the presence of software aging in applications automatically generated by LLMs. Through a controlled experimental setup, four LLM-generated applications were subjected to long-duration workloads, and their behavior was monitored over time using memory usage, response time, and throughput as primary indicators. The results show clear signs of aging-related degradation in all tested systems. In particular, memory consumption exhibited consistent upward trends, and response times fluctuated or increased progressively in several applications. Statistical analysis confirmed the significance of these patterns, supporting the hypothesis that automatically generated software is susceptible to aging. The intensity and nature of the observed symptoms varied among applications, with more complex tasks such as image processing showing faster degradation. These findings suggest that LLM-generated systems should not be assumed stable during prolonged execution periods, highlighting the need for continuous monitoring and corrective strategies. While aligning with traditional software aging behaviors, the results also show sensitivity to workload types and generation choices.

Future work will explore aging mitigation strategies, such as automated rejuvenation, and extend the experimental framework to include a broader set of generation tools, code bases, and runtime conditions. In addition, we plan to investigate the role of specific LLM prompt patterns and code generation strategies in influencing the long-term stability of the resulting applications.

\bibliographystyle{IEEEtran}
\bibliography{full_bibliography}

\begin{thebibliography}{10}
\providecommand{\url}[1]{#1}
\csname url@samestyle\endcsname
\providecommand{\newblock}{\relax}
\providecommand{\bibinfo}[2]{#2}
\providecommand{\BIBentrySTDinterwordspacing}{\spaceskip=0pt\relax}
\providecommand{\BIBentryALTinterwordstretchfactor}{4}
\providecommand{\BIBentryALTinterwordspacing}{\spaceskip=\fontdimen2\font plus
\BIBentryALTinterwordstretchfactor\fontdimen3\font minus
  \fontdimen4\font\relax}
\providecommand{\BIBforeignlanguage}[2]{{%
\expandafter\ifx\csname l@#1\endcsname\relax
\typeout{** WARNING: IEEEtran.bst: No hyphenation pattern has been}%
\typeout{** loaded for the language `#1'. Using the pattern for}%
\typeout{** the default language instead.}%
\else
\language=\csname l@#1\endcsname
\fi
#2}}
\providecommand{\BIBdecl}{\relax}
\BIBdecl

\bibitem{2024Accelerating}
Anupriya, P.~Jain, L.~Goel, R.~Sharma, S.~Jasola, and A.~Ali, ``Accelerating
  software development with artificial intelligence,'' \emph{2024 International
  Conference on Artificial Intelligence and Emerging Technology (Global AI
  Summit)}, pp. 727--732, 2024.

\bibitem{chauhan2025llm}
S.~Chauhan, Z.~Rasheed, A.~M. Sami, Z.~Zhang, J.~Rasku, K.-K. Kemell, and
  P.~Abrahamsson, ``Llm-generated microservice implementations from restful api
  definitions,'' \emph{arXiv preprint arXiv:2502.09766}, 2025.

\bibitem{pietrantuono2022empirical}
R.~Pietrantuono, D.~Cotroneo, E.~Andrade, and F.~Machida, ``An empirical study
  on software aging of long-running object detection algorithms,'' in
  \emph{2022 IEEE 22nd International Conference on Software Quality,
  Reliability and Security (QRS)}.\hskip 1em plus 0.5em minus 0.4em\relax IEEE,
  2022, pp. 1091--1102.

\bibitem{nascimento2024comparison}
M.~G. Nascimento, R.~J. Moura, F.~Machida, and E.~Andrade, ``Comparison of
  machine learning algorithms for detecting software aging in sql server,'' in
  \emph{Proceedings of the 13th Latin-American Symposium on Dependable and
  Secure Computing}, 2024, pp. 159--164.

\bibitem{couto2024comparative}
H.~Couto, F.~Machida, G.~Callou, and E.~Andrade, ``A comparative analysis of
  software aging in relational database system environments,'' \emph{IEEE
  Transactions on Emerging Topics in Computing}, 2024.

\bibitem{bolt2024}
B.~A. Team, ``Bolt: Prompt-based code generation platform,''
  \url{https://boltplatform.ai}, 2024, accessed: 2025-07-14.

\bibitem{Vero2025BaxBench:}
M.~Vero, N.~Mündler, V.~Chibotaru, V.~Raychev, M.~Baader, N.~Jovanovi'c,
  J.~He, and M.~T. Vechev, ``Baxbench: Can llms generate correct and secure
  backends?'' \emph{ArXiv}, vol. abs/2502.11844, 2025.

\bibitem{Lyu2024Automatic}
M.~R. Lyu, B.~Ray, A.~Roychoudhury, S.~H. Tan, and P.~Thongtanunam, ``Automatic
  programming: Large language models and beyond,'' \emph{ACM Transactions on
  Software Engineering and Methodology}, vol.~34, pp. 1 -- 33, 2024.

\bibitem{yang2024automated}
Z.~Yang, F.~Liu, Z.~Yu, J.~Li, and outros, ``Exploring and unleashing the power
  of large language models in automated code translation,'' \emph{arXiv
  preprint arXiv:2404.14646}, 2024.

\bibitem{lu2025webgenbenchevaluatingllmsgenerating}
\BIBentryALTinterwordspacing
Z.~Lu, Y.~Yang, H.~Ren, H.~Hou, H.~Xiao, K.~Wang, W.~Shi, A.~Zhou, M.~Zhan, and
  H.~Li, ``Webgen-bench: Evaluating llms on generating interactive and
  functional websites from scratch,'' 2025. [Online]. Available:
  \url{https://arxiv.org/abs/2505.03733}
\BIBentrySTDinterwordspacing

\bibitem{Balzer1985A}
R.~Balzer, ``A 15 year perspective on automatic programming,'' \emph{IEEE
  Transactions on Software Engineering}, vol. SE-11, pp. 1257--1268, 1985.

\bibitem{Grebenshchikov2012Synthesizing}
S.~Grebenshchikov, N.~P. Lopes, C.~Popeea, and A.~Rybalchenko, ``Synthesizing
  software verifiers from proof rules,'' \emph{Proceedings of the 33rd ACM
  SIGPLAN Conference on Programming Language Design and Implementation}, 2012.

\bibitem{Fan2022Improving}
Z.~Fan, X.~Gao, A.~Roychoudhury, and S.~H. Tan, ``Improving automatically
  generated code from codex via automated program repair,'' \emph{ArXiv}, vol.
  abs/2205.10583, 2022.

\bibitem{Grottke2006Analysis}
M.~Grottke, L.~Li, K.~Vaidyanathan, and K.~S. Trivedi, ``Analysis of software
  aging in a web server,'' \emph{IEEE Transactions on Reliability}, vol.~55,
  pp. 411--420, 2006.

\bibitem{Matias2010Accelerated}
R.~Matias, P.~A. Barbetta, K.~S. Trivedi, and P.~J. de~Freitas~Filho,
  ``Accelerated degradation tests applied to software aging experiments,''
  \emph{IEEE Transactions on Reliability}, vol.~59, pp. 102--114, 2010.

\bibitem{Machida2017Lifetime}
F.~Machida, J.~Xiang, K.~Tadano, and Y.~Maeno, ``Lifetime extension of software
  execution subject to aging,'' \emph{IEEE Transactions on Reliability},
  vol.~66, pp. 123--134, 2017.

\bibitem{SqlServerAgingNascimento2024}
\BIBentryALTinterwordspacing
M.~G. Nascimento, R.~J. Moura, F.~Machida, and E.~Andrade, ``Comparison of
  machine learning algorithms for detecting software aging in sql server,'' in
  \emph{Proceedings of the 13th Latin-American Symposium on Dependable and
  Secure Computing}, ser. LADC '24.\hskip 1em plus 0.5em minus 0.4em\relax New
  York, NY, USA: Association for Computing Machinery, 2024, p. 159–164.
  [Online]. Available: \url{https://doi.org/10.1145/3697090.3699798}
\BIBentrySTDinterwordspacing

\bibitem{security2024passwords}
Security.org, ``2024 password manager industry report and statistics,''
  \url{https://security.org/password-manager-statistics}, 2024, accessed:
  2025-07-17.

\bibitem{uptime2024monitoring}
Uptime.com, ``Website monitoring trends in 2024,''
  \url{https://uptime.com/blog/website-monitoring-trends-in-2024}, 2024,
  accessed: 2025-07-17.

\bibitem{gaibor2024boostlet}
E.~Gaibor, S.~Varade, R.~Deshmukh, T.~Meyer, M.~Geshvadi, S.~Kim, V.~S.
  Narayanappa, and D.~Haehn, ``{Boostlet}.js: Image processing plugins for the
  web via javascript injection,'' \emph{arXiv preprint arXiv:2405.07868}, 2024.

\bibitem{mann1945nonparametric}
H.~B. Mann, ``Nonparametric tests against trend,'' \emph{Econometrica: Journal
  of the econometric society}, pp. 245--259, 1945.

\bibitem{sen1968estimates}
P.~K. Sen, ``Estimates of the regression coefficient based on kendall's tau,''
  \emph{Journal of the American statistical association}, vol.~63, no. 324, pp.
  1379--1389, 1968.

\bibitem{machida2013effectiveness}
F.~Machida, A.~Andrzejak, R.~Matias, and E.~Vicente, ``On the effectiveness of
  mann-kendall test for detection of software aging,'' in \emph{2013 IEEE
  International Symposium on Software Reliability Engineering Workshops
  (ISSREW)}.\hskip 1em plus 0.5em minus 0.4em\relax IEEE, 2013, pp. 269--274.

\end{thebibliography}

\end{document}